\documentclass[runningheads]{llncs}
\usepackage{xcolor}
\usepackage{graphicx}
\usepackage{amsfonts}
\graphicspath{ {./images/} }
\usepackage{array,multirow}
\usepackage{siunitx}

\begin{document}
\title{Deep Generative Models to Simulate 2D Patient-Specific Ultrasound Images in Real Time}
\titlerunning{Deep Generative US simulation}
%

\author{Cesare Magnetti\inst{1} 
\thanks{This work was supported by the Wellcome Trust IEH Award  [102431]. The authors acknowledge financial support from the Department of Health via the National Institute for Health Research (NIHR) comprehensive Biomedical Research Centre award to Guy's \& St Thomas' NHS Foundation Trust in partnership with King's College London and King's College Hospital NHS Foundation Trust.}%
\and
Veronika Zimmer\inst{1}\orcidID{0000-0002-5093-5854} \and
Nooshin Ghavami\inst{1} \and 
Emily Skelton\inst{1} \and 
Jacqueline Matthew\inst{1} \and 
Karen Lloyd\inst{2} \and 
Jo Hajnal\inst{1}\orcidID{0000-0002-2690-5495} \and
Julia A. Schnabel\inst{1}\orcidID{0000-0001-6107-3009} \and
Alberto Gomez\inst{1}\orcidID{0000-0002-7897-7589}}
%
%

\institute{King's College London, London WC2R 2LS, UK.\\ 
School of Biomedical Engineering and Imaging Sciences\\
\url{https://www.kcl.ac.uk/bmeis} \and
Epsom and St Helier University NHS Trust}
\maketitle              
\begin{abstract}

We present a computational method for real-time, patient-specific simulation of 2D ultrasound (US) images. The method uses a large number of tracked ultrasound images to learn a function that maps position and orientation of the transducer to ultrasound images. This is a first step towards realistic patient-specific simulations that will enable improved training and retrospective examination of complex cases. Our models can simulate a 2D image in under 4ms (well within real-time constraints), and produce simulated images that preserve the content (anatomical structures and artefacts) of real ultrasound images. 

\end{abstract}
\section{Introduction}


US imaging is an inexpensive, portable and safe imaging technique. However, it requires a high level of expertise and dexterity from the sonographer to correctly operate the transducer and acquire the standard views. Such expertise is normally acquired through training on patients, which is costly and requires supervision of an expert. This is particularly important in screening clinics, such as fetal screening, because the time allocated for each examination is relatively short and there is a very large patient throughput. The use of ultrasound simulators can be used to support training, however the utility of simulators highly depends on how realistic the resulting images are. Available simulators are not realistic enough to be indistinguishable from real examinations, and particularly lack of real-like non-linear artefacts, variability across patients, fetal motion, etc. 

In this paper, we introduce a novel framework towards real-like simulated ultrasound images. Our proposed method learns the relation between the transducer position (from a tracker) and the resulting images, in a patient-wise fashion. The main contribution of this paper is three fold: first, we introduce a patient-wise learning framework to simulate patient specific images. Second, we compare two different architectures that implement this framework in presence of gaps in the training data. And third, we demonstrate the simulation capability both in phantom and patient data.

\section{Related Work}
Current ultrasound simulators can be classified in three categories ~\cite{bib_1}: interpolative simulators, generative image-based simulators, and generative model-based simulators, as follows:
\begin{itemize}
    \item \textbf{Interpolative simulators:} simulated images are interpolated from previously recorded 3D volumes ~\cite{bib_2,bib_3,bib_4}. These methods can be very fast (can be operated in real time),  but resulting images do not look realistic when the slices are oblique to the volume, mainly because of view dependency of ultrasound image features. Indeed, view dependent artefacts (i.e. shadows, reverberation, etc.) and motion are difficult to simulate with these methods~\cite{bib_5}.
    \item \textbf{Generative image-based simulators:} simulated images are created from existing images obtained from other modalities ~\cite{bib_6}, such as MRI or CT; the output is heavily dependent on the simulation method. Images are normally not very realistic because MRI or CT images are maps of different physical properties ~\cite{bib_7}, and some structures that are visible in ultrasound are just not captured with MRI or CT.
    \item \textbf{Generative model-based simulators:} this category includes two classes: first, physics-based simulators, where a physics simulator is used to produce ultrasound images from a virtual tissue model of anatomy with ultrasound-relevant properties (elasticity, density, etc)  ~\cite{bib_8,bib_9,bib_10,bib_11}. Non-linear modelling of ultrasound propagation through medium is a computationally expensive process and these methods need to trade-off between accuracy and real-timeness, hence are typically not suitable for interactive simulations. Second, ultrasound models learnt from data, a very new field that was recently enabled by the advent of deep learning. The only related work that, to the best of our knowledge has been published is \cite{Hu2017}, where a generative-adversarial model is used to simulate fetal images from a phantom, which is rather homogeneous and does not produce artefacts such as shadows, mirroring and reverberations typical from patient data. In this work, authors use a calibrated coordinate grid as input so accuracy of the simulation depends on the accuracy of this grid, which in turn depends on tracker calibration. 
\end{itemize}

None of the works described above has shown realistic simulations of ultrasound images that fully produce non-linear artefacts and image features such as those found in patient images. In this paper we propose a novel computational framework to build such simulators and demonstrate real-time performance both in phantom and patient data. The resulting simulations are patient-specific, which can enable generating wide range of specific (maybe rare) cases to train on, constituting an invaluable asset for the ultrasound clinic. 


\section{Methods} \label{methods}
We propose two generative Convolutional Neural Networks (CNN) models to produce simulated ultrasound images: first, a decoder  (Fig. \ref{fig:decoder}) to directly map tracking data to images, and second an autoencoder (Fig. \ref{fig:autoencoder}) where the latent space is enforced to represent the tracking information. We also study a third variant which is a decoder where the weights have been pre-trained using an autoencoder. In all cases, we assume that variation between images in the training set depends exclusively on the tracker information (i.e. in this paper we assume no fetal motion). In the following we describe these three architectures.

\subsection{Deep Convolutional Decoder Model}
\label{Vanilla Decoder}

\begin{figure}[!htb]
    \centering
    \includegraphics[width = \linewidth]{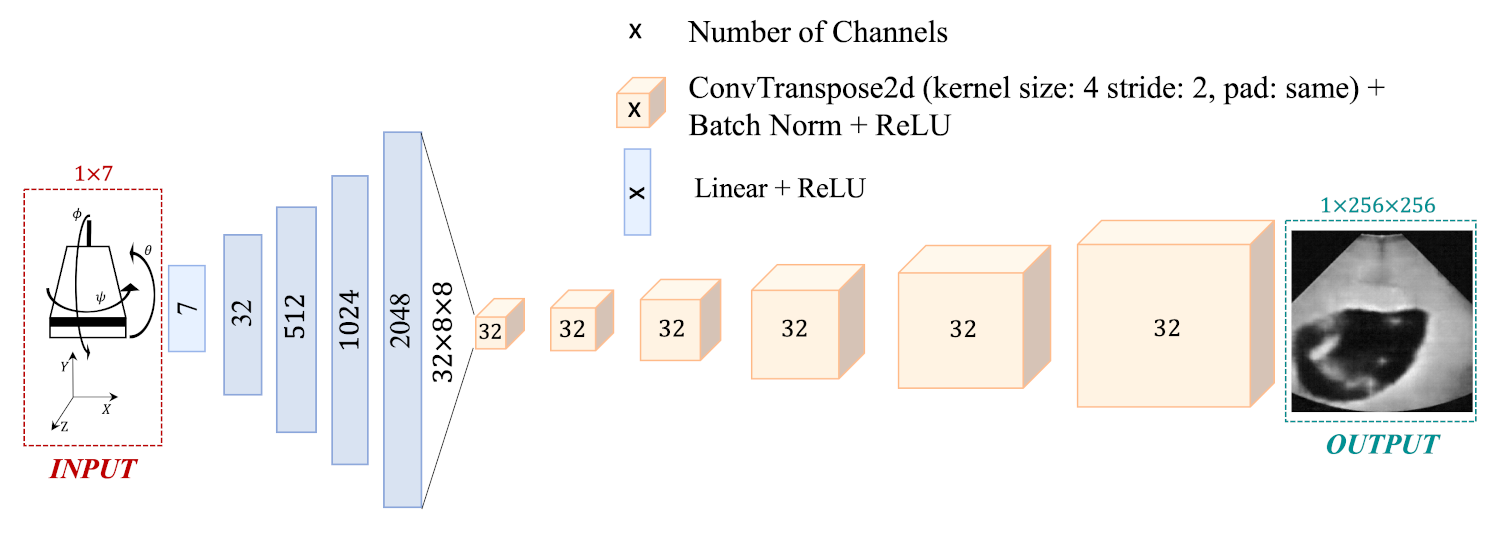}
    \caption{Proposed Deep Convolutional Decoder architecture. The model has five fully connected (FC) layers coupled with ReLU activation functions, followed by seven convolutional layers coupled with ReLU and Batch Normalization. 
    All convolutional layers except the last one have a kernel size of 4 and a stride of 2 and have been padded to achieve a size expansion of 2. The last convolutional layer has kernel size 1 and stride 1 to condense the 32-channel information into  1 channel.}
    \label{fig:decoder}
\end{figure}

We aim at producing simulated ultrasound images ($I_{OUT} \in \mathbb{R}^{N_x \times N_y}$,  with $N_x$ = $N_y$ = 256 pixels) from user-provided 7D tracking data $T_{IN} \in \mathbb{R}^{7}$ representing the transducer pose (4D quaternion + 3D position). We propose a decoder CNN that taking a 7D vector and yielding a 2D image, hence implements the function $I_{OUT} = Simulate(T_{IN})$, with the architecture shown in Fig. \ref{fig:decoder}: 5 Fully Connected (FC) layers followed by a Rectified Linear Unit (ReLU) activation function \cite{nair2010rectified}, that implement a highly non-linear mapping of the 7-D input vector to a rough, low resolution representation of the simulated image, followed by 7 convolutional layers to 
generate a high resolution output image. The parameters of the architecture are provided in Fig. ~\ref{fig:decoder}.

The number of layers of each type was chosen by initially starting with a very deep network (30 layers) and progressively reducing depth until there was no over-fit. At that point, the network was trained with $\pm 4$ layers in the FC and the CNN parts and the number of layers that provided lowest validation 
loss was selected. The number of channels and size of kernels were  empirically chosen trying to minimise the number of parameters while maintaining performance. Two unpooling methods were tested according to~\cite{unpooling}; using strided transpose convolutions worked better than using an upsampling interpolation followed by a convolution and, interestingly, it produced minimal checkerboard-like patterns. The first convolutional feature map with size 4x4 and 8x8 were tested and the latter proved to give better results.

Given a training set with pairs $\{I_{IN},T_{IN}\}$ of real ultrasound images and the transducer poses, the decoder was trained to minimise the loss $\mathcal{L}_{decoder} = MSE(I_{IN},Simulate(T_{IN}))$, where MSE stands for Mean Squared Error. Further details on datasets and training are provided in Sec. \ref{sec:materials_and_experiments}.




\subsection{Deep Convolutional Autoencoder Model} 
\label{Autoencoder}

\begin{figure}[!htb]
    \centering
    \includegraphics[width = \linewidth]{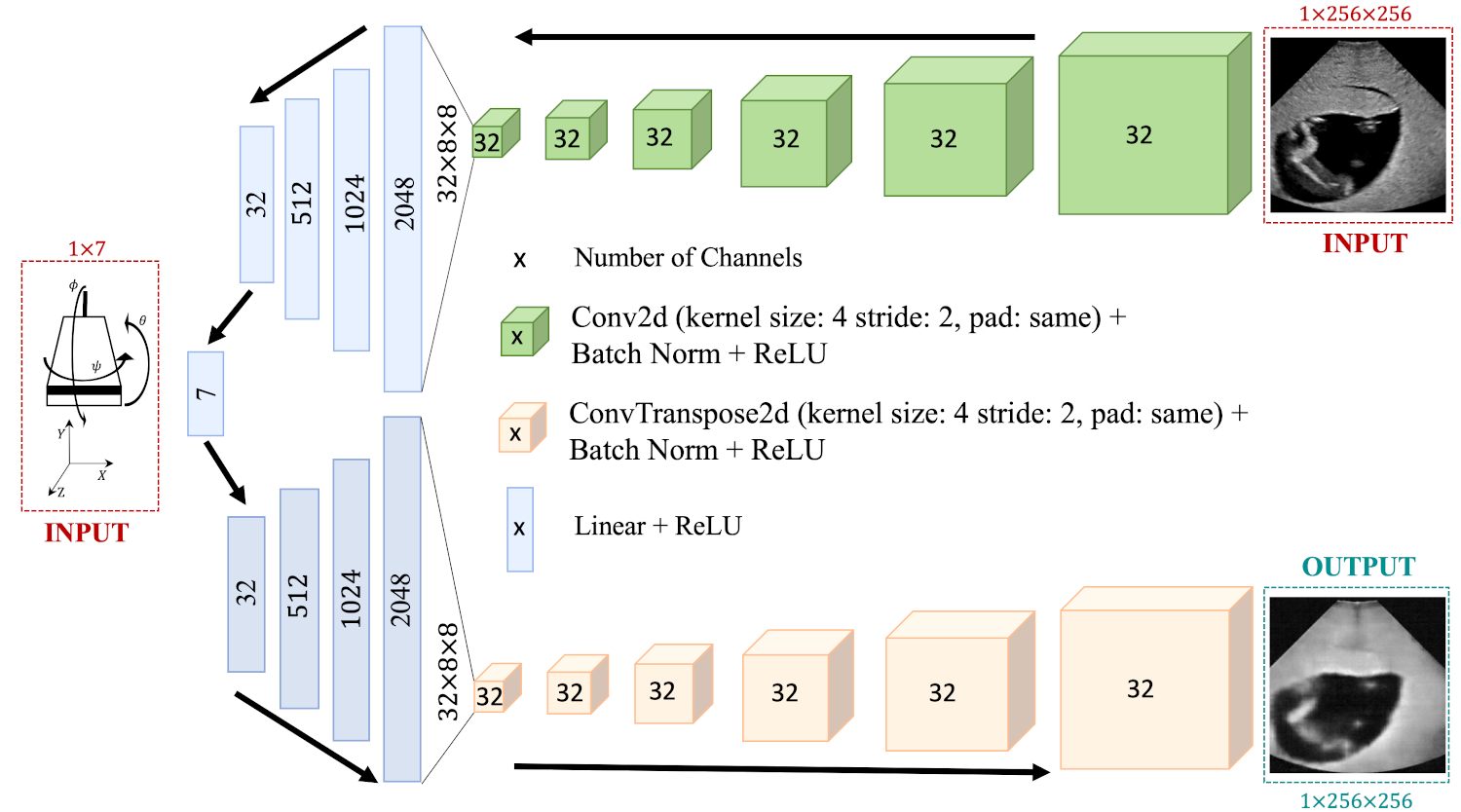}
    \caption{Autoencoder architecture, with the same parameters (kernel size etc.) as the decoder architecture. All convolutional layers are coupled with ReLU and Batch Normalization and all Linear layers are coupled with ReLU. The encoding convolutional layers are standard convolutions, while the decoding layers are transpose convolutions. 
    }
    \label{fig:autoencoder}
\end{figure}

Aiming at improving the model's ability to assimilate image content and to interpolate between training samples, we investigate a multi-input convolutional autoencoder~\cite{autoencoder1,autoencoder2}. This autoencoder architecture consists of an encoder-decoder that mirrors the model described in Sec. \ref{Vanilla Decoder}, 
and is illustrated in Fig. \ref{fig:autoencoder}. 

To implement the multi-input autoencoder, we restrict the latent space $Z$ to be of dimension 7 and enforce disentanglement of the latent space into the components of the tracker data by adding a term in the loss function, as follows. We define the term $\mathcal{L}_{tracker} =  MSE(T_{IN},Z)$, and  with the usual autoencoder loss $\mathcal{L}_{autoencoder} = MSE(I_{IN},Autoencode(I_{IN}))$, the total loss function for the proposed model is:
\begin{equation}
    \mathcal{L}_{multi-input} =  \mathcal{L}_{autoencoder} + K \mathcal{L}_{tracker}
\end{equation}
where $K$ is a non-negative scalar. We empirically found that $K=1$ gave the best results. Other aspects of the training process are provided in Sec. \ref{sec:materials_and_experiments}. At inference time only the decoder part of the network is used, hence the only input during inference is the tracking data, as for the decoder in Sec. ~\ref{Vanilla Decoder}. The number of parameters doubled with respect to the decoder, however for the purpose of inter-model comparison we have left the number of layers untouched.

\subsection{Deep Decoder Model with Pre-trained Weights} \label{pretrained}

The decoder module of the autoencoder is identical to the decoder-only model, hence we propose to pre-train the autoencoder model with the usual $\mathcal{L}_{autoencoder}$ loss (i.e. without any tracking data), and then use the resulting weights as initialization for the decoder model, which is in turn re-trained as described in Sec. ~\ref{Vanilla Decoder}. We hypothesize that this combines the representation learning power of the autoencoder model (and therefore better interpolation capabilities) with a simpler (fewer parameter) model. 

\section{Materials and Experiments}
\label{sec:materials_and_experiments}
In this section all the experiments conducted to test the architectures described in Sec. \ref{methods} will be explained, along with all the details needed to replicate them.

\subsection{Materials} \label{materials}


We carried out experiments both using a phantom (Kyotokagu Space-fan CT) and data from a fetal patient (24 weeks GA). The phantom consisted of 10697 tracked 2D ultrasound images acquired in a single session, plus 11288 untracked 2D ultrasound images that were additionally used for pre-training as described in Sec. \ref{pretrained}. The patient dataset consisted of 15819 tracked 2D ultrasound images, from a single patient during a single scanning session. The ultrasound system used for both datasets was a Philips EPIQ V7 with a X6-1 transducer, with sector width and depth chosen to acquire at 25 Hz (typical in clinical settings). Tracking was done using a NDI Aurora electromagnetic tracker. Images and tracking information were recorded continuously using in-house software. 
Data was split into training and validation sets with a 95\% to 5\% proportion. 
All models were implemented using Pytorch (PyTorch version: 1.1.0, CUDA 8) using a NVidia Quadro M4000 GPU.

Images were pre-processed as follows: first, images were resampled to 0.5mm isotropic resolution and then cropped around the centre to 256x256 pixels using bilinear interpolation. Then, image intensities were rescaled to the interval $(0,1)$ by dividing by 255 (as opposed to using the dynamic range, which may cause abnormally bright images and may alter the user-defined gain settings). To ensure balanced loss terms, the tracking data was scaled dividing \textbf{x}, \textbf{y}, \textbf{z} coordinates by the maximum values  of the NDI tracker, respectively: 250, 250 and 500 mm. Quaternion representation of angles is already normalised to unit vector.

The parameters used for training  all models were: batch size 64 (training) and 16 (validation) 
, dataset randomly shuffled for training, max epochs 200, 40 pre-training epochs, Adam optimizer with a learning rate of 0.0002.

\subsection{Experiments} \label{experiments}
We carried out four types of experiments: first, quantitative quality measurements commonly used to assess image quality. Second, qualitative evaluation of simulation accuracy by a survey to human observers. Third, the impact of low sample-density areas (i.e. holes) in the training set. And fourth, we measure the inference time for each model. 

All experiments are carried out using phantom and patient data except for the third experiment (carried out with phantom data only, where we could obtain as many samples as desired). Details of the experiments are provided next.

\subsubsection{Quantitative Measurement of Image Quality}
\label{sec:materials_and_experiments-1}
We use three quantitative measures: MSE Loss between the original and simulated image; Structural Similarity (SSIM); and Peak Signal to Noise Ratio (PSNR) ~\cite{SSIM_PSNR}. These measurements are widely used to asses quality of ultrasound images and are defined as follows:
\begin{equation}
    \begin{array}{c}
         MSE(I_{IN}, I_{sim}) = \frac{1}{N_{pixels}} \sum_i |I_{IN}(i) - I_{sim}(i)|^2\\\\
         SSIM(I_{IN}, I_{sim}) =  \frac{(2\mu_{I_{IN}}\mu_{I_{sim}} + c_{1})(2\sigma_{I_{IN},I_{sim} + c_{2})}}{(\mu^{2}_{I_{IN}} + \mu^{2}_{I_{sim}} + c_{1})(\sigma^{2}_{I_{IN}} + \sigma^{2}_{I_{sim}} + c_{2})}\\\\
         PSNR(I_{sim}) = 10\log_{10}(\frac{\max^{2}(I_{sim})}{MSE(I_{IN}, I_{sim})})\\
    \end{array}
\end{equation}
where $I_{IN}$ and $I_{sim}$ are the original and the simulated images respectively and $N_{pixels}$ is the total number of pixels in the image. In the second equation, $\mu_{I_{IN}}$ and $\mu_{I_{sim}}$ are the mean values of the original and simulated images and $\sigma_{I_{IN}}$ and $\sigma_{I_{sim}}$ are the standard deviations of the original and simulated images. Lastly, $c_{1} = (k_{1}L)^2$ and $c_{2} = (k_{2}L)^2$ are two variables to stabilise the division with weak denominator, where L is the dynamic range of pixel intensities. $k_{1}$ and $k_{2}$ are scalar constants with default value $k_{1} = 0.01$ and $k_{2} = 0.03$.

\subsubsection{Qualitative Measurement of Image Quality} 
\label{sec:materials_and_experiments-2}
The quantitative measures from previous section need to be complemented by user-rated quality measurements, for which we conducted a survey. We asked six experts specialists (3 fetal sonographers, 3 imaging with expertise in ultrasound) to choose the best quality image from three simulated images corresponding to the proposed models, randomly shuffled. The simulation was carried out at a transducer pose from the validation set and the corresponding real ultrasound image was shown for reference. This was done by 6 users for 100 sets of images each. We report the frequency at which each model was chosen to produce the best image. In the cases where there was a max-vote tie between models, then all were considered best. As a result, the sum of the frequencies over all models may exceed 100\%.

\subsubsection{Impact of Low Density of Samples in the Training Set}
\label{holes}
\begin{figure}[!htb]
    \centering
    \includegraphics[width = 1\linewidth]{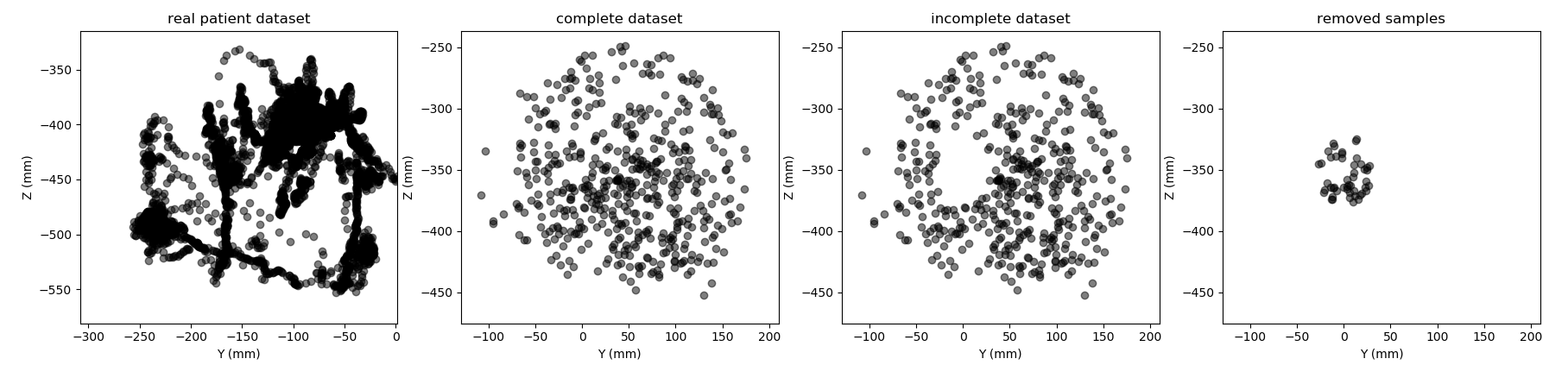}
    \caption{Training data distribution over the half ellipsoidal surface of the fetal phantom, projected onto the `bed plane' for ease of visualization. Data has been randomly downsampled to see individual data points. Each point represents the location of the transducer on the surface of the phantom for an acquired image. From left to right: the complete real-patient dataset, the complete phatom dataset, phantom dataset with a spherical region removed (synthetically created hole with radius=30mm), data removed from the hole. The removed points represented a 12.2\% of the total. Patient data was shown as a proof of the existence of low density of samples regions.}
    \label{fig:incomplete_dataset}
\end{figure}

Some regions of the fetal anatomy will be sampled more densely, because sonographers focus on specific regions throughout the exam. As such, we are interested in analysing the impact of regions with low density of training samples, i.e. holes in the training set, in the simulated images inside and outside the hole, for each model. To this end, we selectively remove training samples within a spherical region arbitrarily located in the phantom dataset, and report the validation loss over the surface of the phantom (shaped as the tummy of a pregnant woman, i.e. half ellipsoid), projected onto a flat plane for easier visualization as illustrated in Fig. \ref{fig:incomplete_dataset}.

We trained all three models with the complete training set (Fig. \ref{fig:incomplete_dataset} second from left) and with the training set minus a small region (Fig. \ref{fig:incomplete_dataset} third from left), and report the performance loss when removing training samples in a localised manner as a relative increase in the validation loss. 

\subsubsection{Inference Time}

We measured the time to simulate an image with all models. To obtain reliable timing results, we measured the time to make 500 inferences and divided by 500, to estimate the per-inference time; and we repeated this process 20 times per model to report average and standard deviation of the inference time.



\section{Results} \label{results}

Table \ref{tab:results} shows the quantitative results obtained from the experiments described in Sec. \ref{sec:materials_and_experiments-1}. Numbers in bold are best for the row, which in all cases corresponds to the decoder architecture. 
\begin{table}[!htb]
\centering
\caption{Quantitative measures of quality for the simulated images: Average MSE Loss (ideal value: 0.0), SSIM Loss (ideal value: 1) and PSNR (in DeciBel (dB), the larger the better). For each model we report the validation (V) and the training (T) losses. The values highlighted in bold indicate the best validation and training results for each row.}
\begin{tabular}{|c|l|c|c|c|c|c|c|}
\hline
& Quality Measure &  \multicolumn{2}{c|}{Decoder}   & \multicolumn{2}{c|}{Autoencoder} & \multicolumn{2}{c|}{Pretrained Decoder} \\
& & V & T & V & T & V & T \\
\hline
\parbox[t]{2mm}{\multirow{3}{*}{\rotatebox[origin=c]{90}{Phant.}}} 
  &Average MSE & \textbf{0.012} & \textbf{0.008} & 0.0240 & 0.0233 & 0.0150 & 0.0146\\
 & Average SSIM & \textbf{0.455} & \textbf{0.480} & 0.453 & 0.456 & 0.422 & 0.425\\
 & Average PSNR & \textbf{19.19} & \textbf{20.85} & 16.23 & 16.50 & 18.12 & 18.42\\
\hline
\parbox[t]{2mm}{\multirow{3}{*}{\rotatebox[origin=c]{90}{Patient}}} 
& Average MSE & \textbf{0.0061} & \textbf{0.0055} & 0.0107 & 0.0101 & 0.0068 & 0.0065\\
& Average SSIM & \textbf{0.656} & \textbf{0.660} & 0.638 & 0.643 & 0.641 & 0.644\\
& Average PSNR & \textbf{22.93} & \textbf{23.81} & 20.10 & 20.90 & 22.19 & 22.76\\
\hline
\end{tabular}
\label{tab:results}
\end{table}

\begin{figure}[!htb]
    \centering
    \includegraphics[width = .7\linewidth]{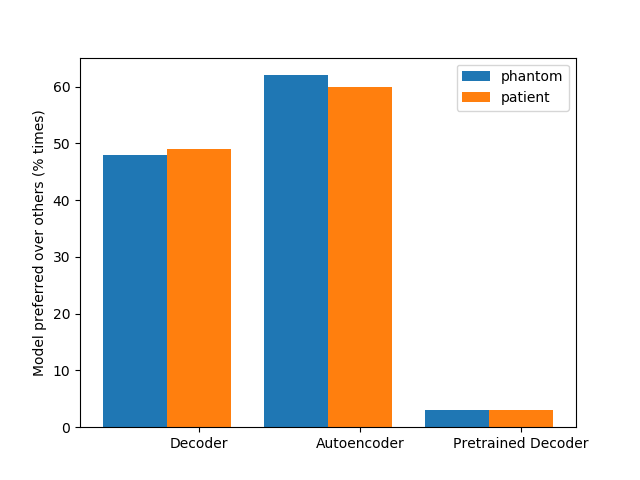}
    \caption{Bar plot showing the proportion of the cases (in \%) where each model was voted best by all raters. When two models tied at having maximum votes, both were considered best hence sum of bars may exceed 100\%.}
    \label{fig:survey results}
\end{figure}

The qualitative results, from the experiment described in Sec. \ref{sec:materials_and_experiments-2}, are illustrated in Fig. \ref{fig:survey results}. The bars show the proportion of the cases (in \%) where each model was voted best by the 6 scorers. It can be observed that results are consistent accross datasets, with the autoencoder being perceived as best closely followed by the decoder. In all cases the pretrained decoder was found to have poor perceptual quality compared to the other two.

\begin{figure}[!htb]
    \centering
    \begin{minipage}{0.02\linewidth}
    \rotatebox[origin=c]{90}{Decoder}
    \end{minipage}
    \begin{minipage}{0.97\linewidth}
    \includegraphics[scale = 2, width = \linewidth]{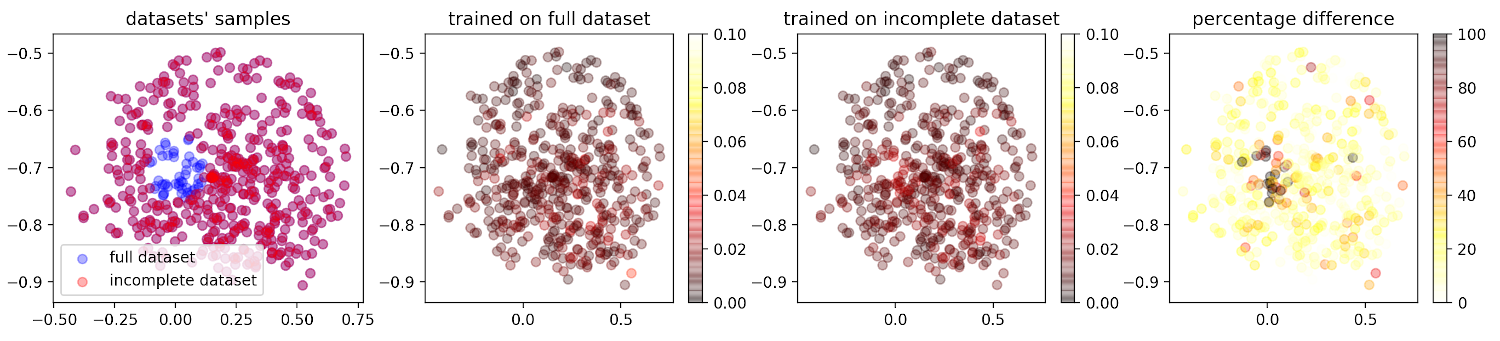}
    \end{minipage}
    \centering
    \begin{minipage}{0.02\linewidth}
    \rotatebox[origin=c]{90}{Autoencoder}
    \end{minipage}
    \begin{minipage}{0.97\linewidth}
    \includegraphics[scale = 2, width = \linewidth]{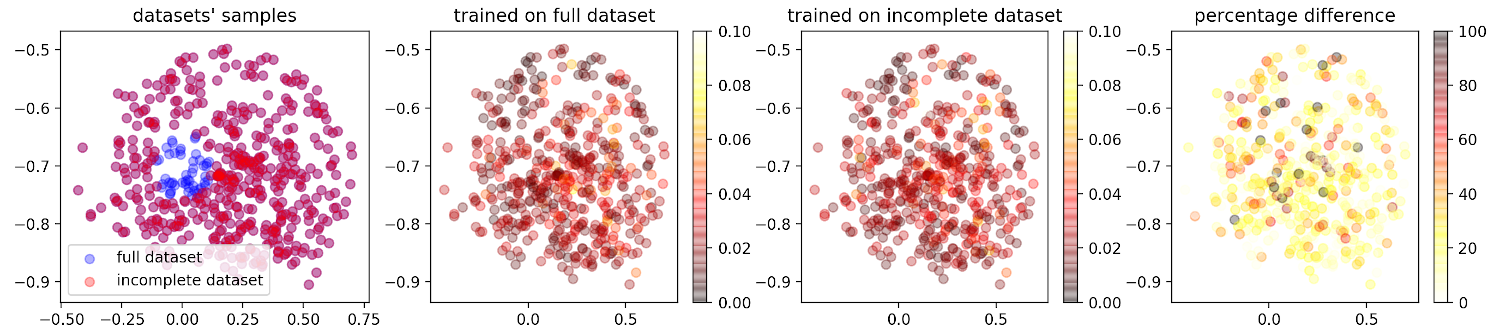}
    \end{minipage}
    \centering
    \begin{minipage}{0.02\linewidth}
    \rotatebox[origin=c]{90}{Pre. Decoder}
    \end{minipage}
    \begin{minipage}{0.97\linewidth}
    \includegraphics[scale = 2, width = 1\linewidth]{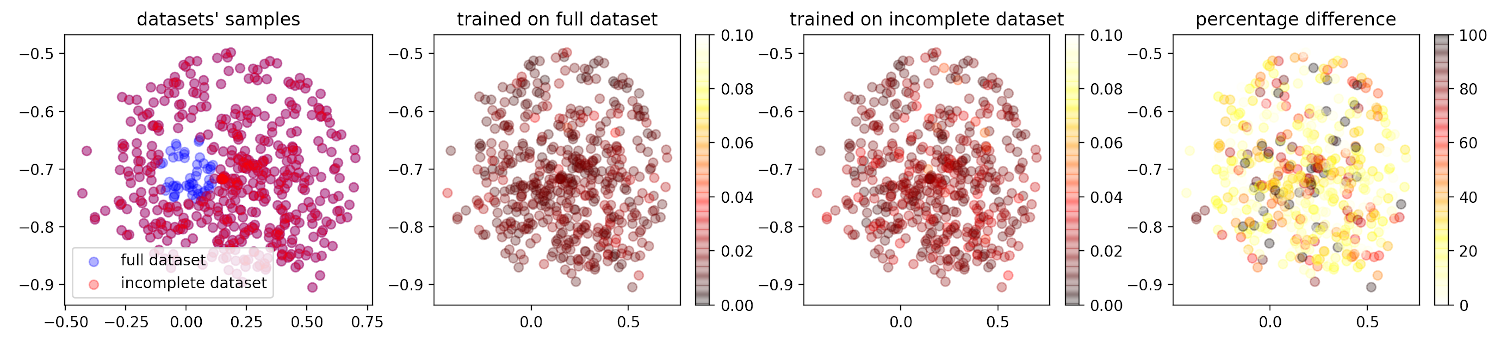}
    \end{minipage}
    \caption{A diagram showing the MSE Loss between real and simulated images for each sample in the validation set. Column 1 is showing the two different datasets used; according to Section \ref{holes} the model was trained on the full dataset, with results depicted in column 2, and on an incomplete dataset, with results shown in column 3. Column 4 shows the percentage difference in the MSE Losses of the two previous scenarios, allowing for easier interpretation. Top row shows results for Architecture \ref{Vanilla Decoder}, middle row shows results for Architecture \ref{Autoencoder}, bottom row shows results for Architecture \ref{pretrained}}
    \label{fig:loss_maps}
\end{figure}

The results of the experiment on low density of training samples within a region, as described in Sec. \ref{holes}, are shown in Fig. \ref{fig:loss_maps}. The first column shows the points left out from the training set in blue. The second column shows the loss using the entire dataset. The third column shows the loss using the training set without the blue region from the first column. The last column shows the relative increase in loss (in \%) when removing a region of the training set. The decoder architecture shows a very localised reduction of performance, while the architectures that use an autoencoder seem to maintain a more homogeneous loss. 

Results on inference time are the following, given in average $\pm$ standard deviation; Decoder:  3.6ms $\pm$  0.53ms. Autoencoder: 3.7ms $\pm$ 0.56ms; and pre-trained Decoder 3.7ms $\pm$ 0.52ms. These values are all largely within real-time constrains, which for 25 Hz requires an inference time of up to 40ms.

\begin{figure}[!htb]
    \centering
    \includegraphics[width = 1\linewidth]{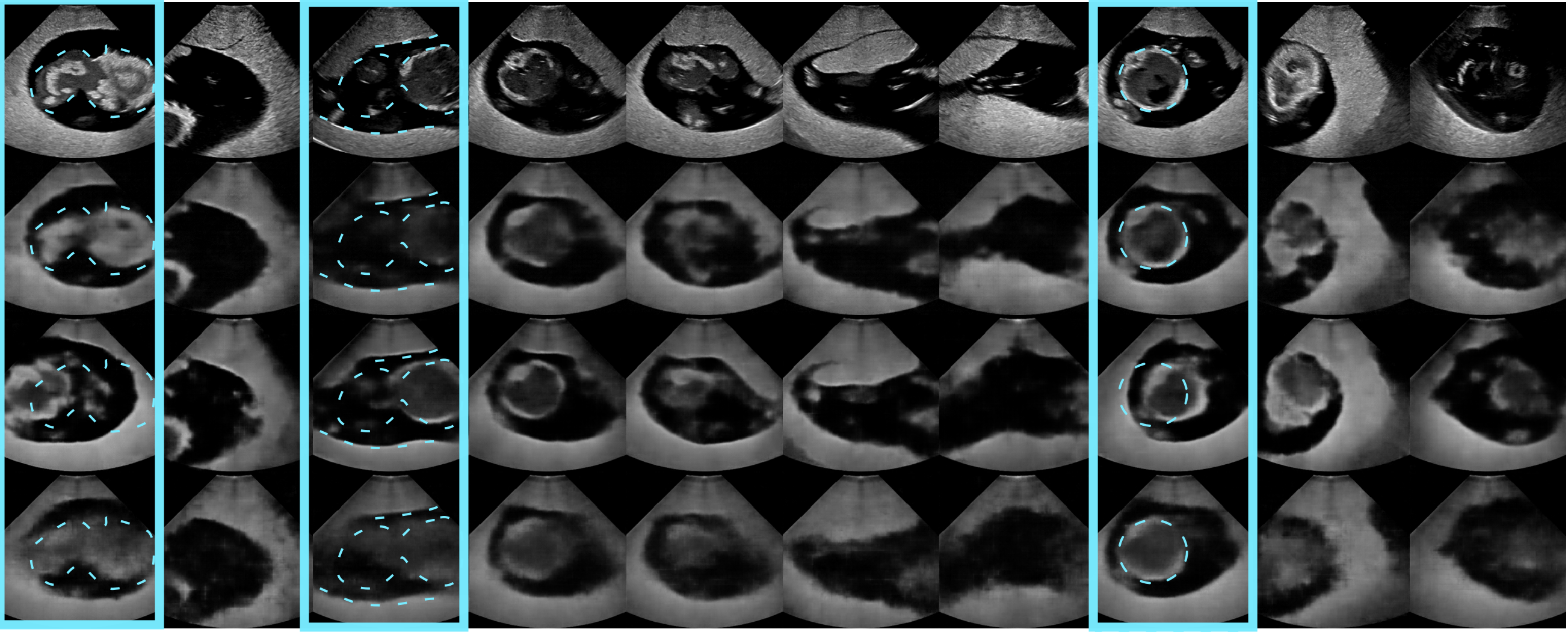}
    \caption{10 examples of simulated images on phantom data for each architecture, using tracking data from the validation set. The top row shows the original images, The second row from the top shows images simulated with the decoder (Sec. \ref{Vanilla Decoder}), the third row shows images simulated with the autoencoder (Sec. \ref{Autoencoder}) and the bottom row shows images simulated by the pre-trained decoder (Sec.  \ref{pretrained}).}
    \label{fig:some_images}
\end{figure}

\begin{figure}[!htb]
    \centering
    \includegraphics[width = 1\linewidth]{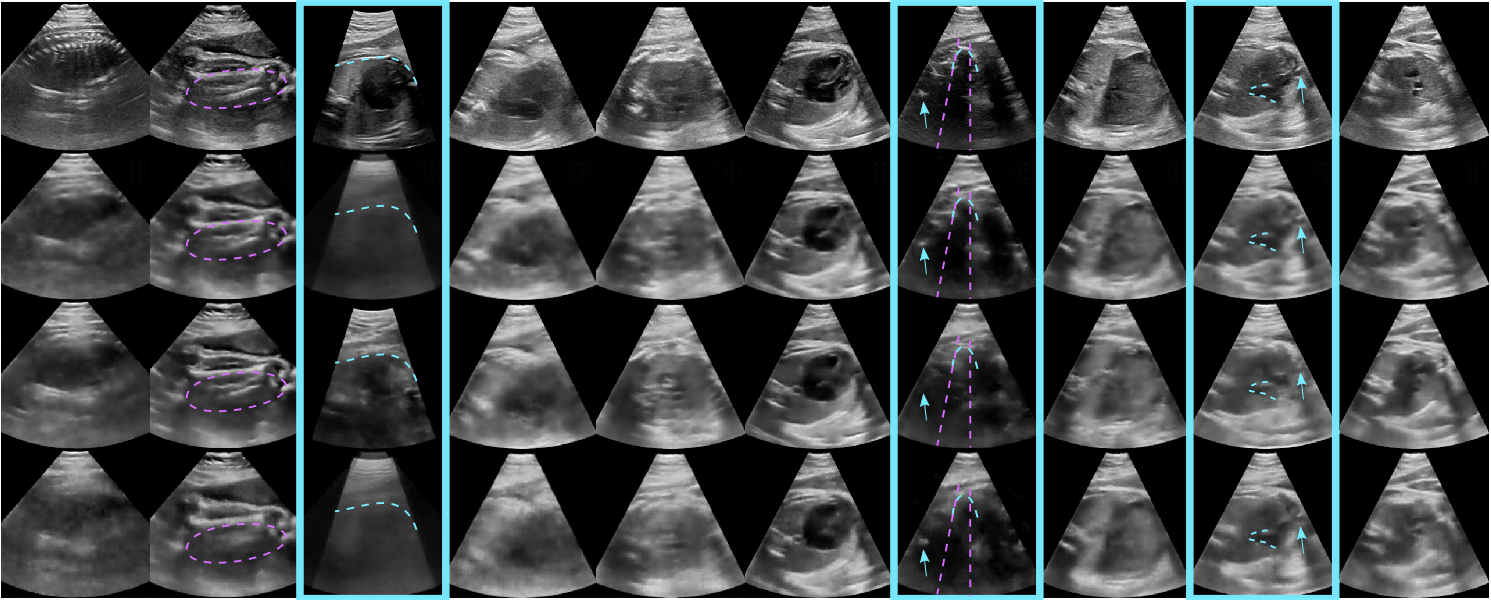}
    \caption{10 examples of simulated images on real patient data for each architecture, using tracking data from the validation set. The top row shows the original images, The second row from the top shows images simulated with the decoder (Sec. \ref{Vanilla Decoder}), the third row shows images simulated with the autoencoder (Sec. \ref{Autoencoder}) and the bottom row shows images simulated by the pre-trained decoder (Sec.  \ref{pretrained}). Dashed lines highlight the difference in performance among the different methods: the blue dashed lines roughly indicate where the main image features are on the original images and how they are reproduced in the simulated images; the purple dashed lines do the same for artefact features (reverberation on the left, and shadow on the right).}
    \label{fig:some_images_pat}
\end{figure}

Finally, for illustrative purposes, we show examples of simulated images produced at 10 randomly picked locations of the validation set. The simulations from the phantom dataset are shown in Fig. \ref{fig:some_images}, and the simulations from the patient dataset are shown in Fig. \ref{fig:some_images_pat}. In both cases, the top row shows the original images, and the rows 2, 3 and 4 show the simulations using the decoder, the autoencoder, and the pre-trained decoder, respectively. Three columns have been highlighted in Fig. \ref{fig:some_images}, with a dashed blue contour indicating the main image features in the original image and where they are shown in the simulated images. These images exemplify cases where the autoencoder architecture sometimes shows high quality images but different from the original image, hence not good for simulation. This suggests overfitting to the training set, probably because this architecture has twice as many parameters as the decoder.
Similarly, three columns have been highlighted in the patient results in Fig. \ref{fig:some_images_pat}, where features of the original images have been indicated with dashed blue lines and arrows. Column three shows an example where the autoencoder is the only architecture able to produce a good simulation. The other two highlighted columns show the same effect found in patient data (high image quality but low simulation quality  with the autoencoder). The purple dashed lines indicate how artefacts (reverberation in column 2 and shadowing in column 7) are simulated. This has not been reflected in the phantom dataset because the phantom does not present large enough differences between the tissue properties and hence does not show any of these artefacts.

\section{Discussion} \label{discussion}

We have presented a method to simulate real-time, patient-specific 2D ultrasound images by training deep convolutional models with data from a single patient consisting of paired tracking data and corresponding images. Our method does not require any tracker calibration (as long as the entire dataset for a patient has been acquired without moving the tracker with respect to the transducer) and, at inference time, takes as input a 7D vector corresponding to the position and orientation of the tracking device. As opposed to the work in \cite{Hu2017}, our framework uses the tracking data directly as input, and as such extending this work to moving organs could be achieved by adding temporal dimensions to the tracking vector. This hypothesis will be verified in future work. 

Of the three proposed architectures, the quantitative results in Table \ref{tab:results} suggest that the decoder model performs best. This is also supported by the visual results show in Figs. \ref{fig:some_images}, \ref{fig:some_images_pat}, where it seems that often the autoencoder shows an image that has good quality but does not correspond to the original image. This may be due to over-fitting on the training set because this model has twice as many parameters as the decoder. The fact that this was less of an issue with patient data, where we used more images, further supports this hypothesis. Interestingly, the user survey shows a slightly better performance of the autoencoder; we believe that this perception could be due to the fact that the quality of the images simulated by the autoencoder is actually higher, which made raters believe that the simulation was better too. 

The quantitative results with the pre-trained decoder are surprisingly poor, and this is reflected visually by images that resemble the decoder simulations in terms of image features but are consistently more blurry. Incorporation of quality improvement mechanisms in the pipeline, e.g. adversarial training such as \cite{Hu2017} will be investigated in future work. We also plan to study interactive simulation to test suitability of the proposed method for training.

Our results also suggest that when there are `holes' in the training set, simulation within these holes may still be possible, although with a loss of accuracy. Our holes were simulated by filtering out the position of the transducer, so further investigation involving the orientation of the transducer is required.

\section{Conclusion}

We presented a novel framework to simulate patient-specific 2D US images in real-time. Of three proposed models, a deep decoder was the best simulator for phantom and fetal patient data, with a simulation time under 4ms per frame.




\bibliographystyle{splncs04}

\end{document}